# Influential Node Ranking in Complex Information Networks Using A Randomized Dynamics-Sensitive Approach

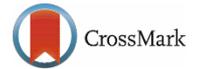

*Ahmad Asgharian Rezaei, Justin Munoz, Mahdi Jalili, Hamid Khayyam*

*School of Engineering, RMIT University, Melbourne, Australia*

**ABSTRACT**

Identifying the most influential nodes in information networks has been the focus of many research studies. This problem has crucial applications in various contexts, such as controlling the propagation of virus or rumours in real-world networks. While existing methods mostly employ local neighbourhood features which heavily rely on the network structure and disregard the underlying diffusion dynamics, in this work we present a randomized sampling algorithm that not only considers the local and global structural features of the network, but also considers the underlying diffusion dynamics and its parameters. The main idea is to compute the influentiality of a node through reachability from that node in a set of random graphs. We use a hyper-graph to capture the reachability from nodes in the original network, and theoretically argue that the hyper-graph can be used to approximate the theoretical influentiality of nodes in the original graph with a factor of $(1-\epsilon)$. The performance of the proposed model is also evaluated empirically by measuring the correlation between the ranking generated by the proposed method and the ground-truth ranking. Our results show that the proposed method substantially outperforms state-of-the-art methods and achieves the highest correlation with the ground-truth ranking, while the generated ranking has a high level of uniqueness and uniformity. Theoretical and practical analysis of the running time of the algorithm also confirms that the proposed method maintains a competitive running time in comparison to state-of-the-art methods.

## 1. Introduction

In recent years, much research has focused on studying the spreading process in complex networks [1, 2], mainly because it can model the spreading phenomena in the real-world [3]. Identifying the influential nodes in a complex network is the first question when studying the spreading process [4], such as disease propagation, rumours diffusion [5], cascading failure, and viral advertising [6]. When vital nodes act in a network, the impact quickly propagates to the whole network. Therefore, developing an accurate method for quantifying the importance of nodes in complex networks is of high theoretical and practical significance [7]. For instance, the development of such methods is crucial for controlling a pandemic through restricting and treating the key nodes in a disease network [8, 9], stopping the diffusion of rumours through controlling the most important nodes in social networks [10], avoiding cascade failure by taking protective measures for the key circuits of the power network [11], accelerating information diffusion, or promoting new products.

When it comes to ranking nodes based on importance in a complex network, degree is the most straightforward and simplest indicator [12]. Although ranking nodes based on their degree has a low computational cost and appears to be an efficient method for large-scale graphs, it suffers from low accuracy. To address the shortcomings of the degree centrality, other centrality measures have been proposed, such as betweenness centrality [13], and closeness [14]. However, not all of these methods are scalable to large-scale networks due to a high computational complexity. The idea of decomposing the network has been developed in methods such as K-Shell [15]. Following K-Shell, other methods including gravity centrality [16] and extended K-Shell sum [17] adopted the decomposition idea to rank nodes based on importance. In a recent work [1], degree centrality is combined with local clustering coefficient and k-shell decomposition to capture both local and global influentiality.

---

\* *Corresponding author.*
E-mail address: a.asghariyan.rezayi@gmail.com (A.A. Rezaei)

In another work, Kitsak et al., [15] showed that nodes in the core part of the network are more likely to be influential. Inspired by this work, Chen et al. [18] proposed the coreness centrality measure. Through K-Shell decomposition, the coreness method assigns many nodes to the same shell. However, nodes of the same shell may have different influentiality. Furthermore, K-Shell may fail to identify influential nodes in core-less networks [19]. In fact, methods based on k-coreness often do not provide satisfactory performance. Recently, improved centrality measures, such as mixed degree centrality (MMD) [20], shell-based ranking and filtering method (SRFM) [21], local structure centrality [22], neighbourhood coreness centrality [23], and H-index [24], have been proposed in which multiple measures were combined to derive a new measure.

In addition to the structural centrality measures that mainly focus on the neighbourhood information of the nodes to quantify influentiality, there are iterative refinement centrality measures, such as eigenvector centrality [25], cumulative nomination [26], PageRank [27], HITS [28] and their variants that take the influentiality of the neighbours of a node into account for computing the influentiality of nodes. A recent method uses structural similarity of the nodes captured by a KL divergence with the PageRank algorithm to improve the classic PageRank method [29]. Often, these methods assign a uniform start score to all nodes in the network and then refine the score through an iterative process to get to a steady state.

One major disadvantage of centrality-based measures is that they do not take the properties of the underlying dynamical process into account. In other words, if the underlying dynamics change, ranking of the nodes based on centrality measures will remain the same. However, it has been shown that even for the same dynamics, under different dynamics parameters, no single centrality measure can always give the most accurate ranking of the nodes [30]. For example, in the Susceptible-Infected-Removed (SIR) dynamics when the spreading rate, $\beta$, is small, the degree centrality can better identify the influentiality of nodes, while the eigenvector centrality performs better when the spreading rate is close to the epidemic threshold $\beta_c$ [31]. To address this issue, alternative methods should also consider the properties and parameters of the target dynamics. We refer to such methods as *dynamics-sensitive methods* [32]. In this regard, Klemm et al. [31] proposed the dynamical influence measure which is the left eigen vector of the largest eigenvalue of the dynamics matrix, assuming a linear approximation of non-linear dynamics like SIR. Following the same idea, Liu et al. [32] used matrix differentiation to calculate the influentiality of nodes. Ide et al. [33] and Bauer [34] also proposed other dynamics sensitive approaches that mainly rely on counting the number paths and walks in a complex network.

In this manuscript, we propose a novel dynamics-sensitive method that employs a randomized sampling algorithm to generate influence paths. We treat each influence path as a hyper-edge of a hyper-graph that has the same nodes as the original network. We theoretically prove and empirically show that the degree of nodes in the hyper-graph is a good indicator of their influentiality. Our results show that the proposed randomized sampling approach yields state-of-the-art results, while the computational complexity is still comparable with existing methods. We also show that the proposed method achieves an unseen stability in performance for different values of parameters of the underlying diffusion dynamics, a behaviour that is absent in the results of the existing dynamics-insensitive models.

## 2. Related Works

In this section, we review some of the most notable existing works about influential node ranking in more details. There are many factors that can affect the ability of a node to propagate its influence throughout the network such as the structure of its neighbourhood, its location in the network, the content and context of the message and many more [35]. Since obtaining contextual information is challenging in many cases, the structural and topology-based features of a network are primarily used for determining the influentiality of nodes in many of the existing methods. This in turn makes these approaches independent of the underlying dynamical processes. The notion of centrality in which a real value is assigned to the nodes in a network provides a measure for quantifying the vitality of the nodes.

The most intuitive centrality measure is the degree centrality, whereby the importance of the nodes is quantified by the number of their neighbours. Although this measure appears to be very simplistic, it can achieve a surprisingly good performance, particularly when the propagation rate is very small. In fact in the latter case, degree centrality outperforms some of the more sophisticated centralities such as the eigenvector centrality [31]. An improved version of degree centrality, LocalRank, was proposed by Chen et al. [18] extending degree centrality by considering the information of the fourth-order neighbours of the nodes. LocalRank algorithm has a lower computational complexity compared to the path-based centralities, while achieving a competitive performance [18]. Inspired by the fact that local interconnectedness has a negative correlation with the size of the propagation [36, 37], Chen et al. [38] proposed ClusterRank which considers the clustering coefficient of a node [39], as well as the number of neighbours. The clustering coefficient is used to incorporate the interconnectedness effect into the new centrality, as a large clustering coefficient is an indicator of the of high interconnectedness and low influence. We use $CR(i)$ to denote the ClusterRank of node $i$. We have:

$$CR(i) = f(c_i) \sum_{j \in \Gamma_i} (k_j^{out} + 1)$$

In the above formulation, $c_i$ is the clustering coeficient of node $i$, $f$ is a function of the cluster coefficient, and $k_j^{out}$ is the out degree of node $j$.

Recently, Kitsak et al. [15] introduced a new family of centrality measures focussing more on the location of the node in the graph as opposed to its degree and neighbourhood structure. The authors argue that if a node is located in the core part of the network, its influence is higher compared to other nodes that are located in the outer parts of the network. The new measure is referred to as coreness and is obtained by applying k-core decomposition in networks [40]. The k-core decomposition process is as follows: given an input graph G, every isolated node with a degree of zero receives a coreness score of 0 and are then removed from the network. Next, all the nodes with a degree of one are removed from the network and will receive a coreness score of 1. Following the removal of nodes with a degree of 1, the degree of the remaining nodes will decrease by 1. Therefore, nodes with a residual degree less than equal to 1 should also be removed from the network. These nodes will also receive a coreness score of 1. The process will continue by removing nodes with a residual degree of 2, and then removing all the nodes with a residual degree less than equal to 2 and so on. The coreness score of nodes is equal to the degree threshold by which nodes are selected for removal. Due to its low computational complexity, the k-shell decomposition method also known as the KS or the k-core method can be used for large-scale networks and in fact has many applications in real networks [7].

The k-core method, however, only focuses on the residual degrees of the nodes and drops all the information in relation to the removed nodes. In contrast, Zeng et al. [20] proposed a new decomposition method called mixed degree decomposition (MMD), in which both the residual degree of the nodes along with the exhausted degree of the nodes are considered when picking the nodes for removal. In fact, in each step, nodes are removed based on the mixed degree which is calculated through the following formulation:

$$k^m = k^r + \lambda k^e$$

where $k^m$ is the mixed degree, $k^r$ is the residual degree, $k^e$ is the exhausted degree, and $\lambda$ is a tenable parameter between 0 and 1.

Another drawback of the k-core method is that it assigns the same coreness score to many nodes with a different degree resulting in a ranking list whereby many nodes have the same rank. To address this issue, Bae and Kim [23] proposed a new measure that considers both the degree and coreness score of a node to calculate a new score. The main idea behind this centrality measure is the assumption that nodes which reside in the central part of the network and have a large number of neighbours are more vital. Based on this idea, the neighbourhood coreness is defined as follows.

$$C_{nc}(v) = \sum_{w \in \Gamma(v)} ks(w)$$

where $\Gamma(v)$ is used to denote the set of neighbours of node $v$, and $ks(w)$ shows the k-shell score of node $w$. By adding the $C_{nc}$ scores of all the neighbours of a node, the $C_{nc+}$ score is defined recursively. We have:

$$C_{nc+}(v) = \sum_{w \in \Gamma(v)} C_{nc}(w)$$

The $C_{nc}$ method improves the traditional KS algorithm, but still shares some disadvantages with the main algorithm. Both the KS method and the method proposed by Bae and Kim [23] disregard the iteration information. In fact, in these algorithms, some nodes may end up having the same coreness score with the coreness score assigned to them at different iterations. To address this issue, Wang et al. [41] proposed a method that also incorporates the iteration information by defining a k-shell iteration factor. For each node, the k-shell iteration factor is calculated using the following formula, where $k$ is the coreness score of the node, $n$ is the iteration in which the score is assigned to the node, and $m$ is the total number of iterations for processing all nodes with coreness score $k$.

$$\delta_u = k.(1 + \frac{n}{m})$$

Using the k-shell iteration factor, the influence capability of a node $u$ is defined as follows:

$$IC_u = \delta_u.d_u + \sum_{v \in \Gamma(u)} \delta_v d_v$$

where $d_u$ is the degree of node $u$, and $\Gamma(u)$ is the set of all neighbours of node $u$.

In a similar approach, Zareie et al. [42] changed the notion of k-shell iteration factor, and introduced shell clustering coefficient where for each node, the Pearson correlation between the shell vector of the target node and all of its neighbours are added together. For this new measure we have:

$$SCC_u = \sum_{v \in \Gamma(u)} (2 - C_{uv}) + (2.\frac{\deg(u)}{\max_{k \in G} \deg(k)} + 1)$$

where $C_{uv}$ is the correlation between the nodes $u$ and $v$. The cluster coefficient ranking measure for a node $u$ is then defined as the sum of the shell clustering coefficient for all of its neighbours.

Although many improvements have been made to the original KS algorithm, all of the KS-based methods still require global topological information of the network, which is not always available in real-world scenarios. We refer to such centrality measures as global centrality measures. In contrast, local centrality measures only require partial information of the network to be computed. H-index [43] is a local centrality measure that only requires the degrees of neighbours of a node. H-index was originally proposed for quantifying the impact of researchers based on the number of received citations but was later extended to measure the influentiality of nodes in social networks. In the extended version of $H$-index, a node $v_i$ has an h-index of h, if it has h neighbours all of which with a degree not less than h [44].

There are also several works that combine the notion of coreness along with other concepts to derive new ranking measures. The MCDE [45] method adds up a factor of coreness with a factor of degree, and a factor of entropy for ranking nodes based on their vitality. In a similar work, Zareie et al. [46] proposed the diversity-strength ranking where entropy of the KS score of nodes over the KS score of their neighbours is the main component of the new centrality.

Furthermore, Liu et al. [32] used a matrix differential function to calculate the spreading influence of nodes in a given network. The matrix differential function in this method also takes influence probability as an input, making the scoring function sensitive to the parameters of the dynamics. The simplified version of this matrix differential function can be expressed as the sum of powers of the adjacency matrix, multiplied by powers of the influence probability. Although this method appears to be dynamics-sensitive, it requires adjacency matrix multiplication which is not practical for large networks. Moreover, our experiments highlight that this method does not show much of dynamic sensitivity when the influence probability changes in an SIR dynamic. In another research, Ide et al. [33] proposed a dynamic-sensitive approach for the Susceptible-Infected- Susceptible (SIS) model, which is argued to be equal to a typical path counting method. Similarly, Bauer and Lizier [34] proposed a method that directly counts the number of possible walks of various length in the SIR and SIS model where the dynamics properties are used to design the decaying faction of path weight. However, since counting the number of paths and walks in complex networks has a high computational cost, the two latter methods are less efficient for large-scale networks.

There are many more developments in centrality measures, but we have only limited the focus of this section to the state-of-the-art methods that have achieved outstanding ranking performance on a number of real and synthetic datasets. In the results section, we have compared the performance of our proposed model with some of the models that we briefly introduced in this section.

## 3. Proposed Method

Let us denote a network by graph $G = (V, E)$, where $V$ represent the nodes and $E \subseteq V \times V$ is the set of undirected edges between nodes. In this graph, if there is an edge $e = (u, v)$ connecting two nodes $u$ and $v$, we refer to $u$ and $v$ as neighbors. The notation $\Gamma(v)$ is used to represent the set of neighbours of a given node $v$. The number of neighbours of a node shows the degree of that node or, $\deg_G(v) = |\Gamma(v)|$.

In this work we study the Susceptible-Infected-Removed (SIR) model [47], as the underlying dynamics that models the spreading process in the network. From now on, the term propagation dynamics, will also act as a reference to SIR. Given graph $G$ and a node $v$ as the infected node (known as the seed node), the SIR process for the case when $\mu = 1$ is as follows:

- At round 1, only node $v$ is in the infected state, and the rest of the nodes are in susceptible state. At this round, node $v$ infects (influences) each of its neighbours with infection (influence) probability $\beta$. Infection probability is a parameter of the SIR model that is considered to be uniform for all nodes in this work.
- At round $t > 1$, all nodes that were at the infected state at the previous round move to the removed state with probability 1. This happens since $\mu = 1$. Nodes at the removed state will no longer influence other nodes or get influenced by others. Moreover, nodes that were at the susceptible state and become infected at round $t - 1$ will move to the infected state at round $t$. These newly infected nodes will now infect each of their neighbours with probability $\beta$.
- The influence propagation stops when no more nodes become infected at the last round.
- At the end of the propagation, the total number of nodes that are in a removed state divided by the network size marks the influentiality of $v$.

Since influence propagation is a stochastic process, we use $I(v)$ to show the expected number of infected nodes in a propagation process. In fact, if we use $Dist_I(v, G, D)$ to show the distribution of the influence-spread of node $v$ in graph $G$ under dynamics $D$, $I(v)$ is a random sample from this distribution. Influential Node Ranking (INR) is to sort the nodes of a network based on their influentiality [46]. The main drawback of the existing approaches for INR is the insensitivity of the influence-spread (influentiality) estimation to the underlying dynamics. Most of the existing approaches only rely on structural properties of $G$ to estimate $I(v)$ and disregard the parameters of the dynamics.

In this work, we introduce a novel dynamics-sensitive algorithm that efficiently estimates the influentiality of nodes while taking account the underlying dynamics parameters. In order to do so, we first show that the influentiality of a node in a given graph $G$ under the SIR model is equivalent to the reachability from that node in random sub-graphs of $G$. By learning from this lemma, we aim to estimate influentiality of nodes in $G$ through computing reachable nodes from them in sub-graphs of $G$. We build a hyper-graph on the nodes of $G$, with hyper-edges that model reachability between the nodes in $G$. The hyper-edges are sampled using a randomized algorithm. We show that if enough random graphs (hyper-edges) are sampled, then the degree of nodes in the hyper-graph will approximate to the influentiality of nodes in the original graph, $G$. We then conclude the theoretical section of the paper by giving a lower bound for the number of random sub-graphs (hyper-edges) to sample.

**Definition 1 (Random Sub-graphs of G: $\beta$-graphs)** Given $G = (V, E)$ as the original graph, a $\beta$-graph is a random graph with a subset of nodes of $G$ as $V_\beta$. Let $E_\beta$ denote edges of the $\beta$-graph where $\beta < 1$, then each edge $e \in E$ will be added to $E_\beta$ with probability $\beta$, and $V_\beta$ only includes nodes that have an active edge in $E_\beta$.

**Lemma 1.** The influentiality of a node $v \in V_{g_\beta}$ is equal to the number of nodes reachable from $v$ on $g_\beta$.

**Proof.** Each edge in $g_\beta$ is picked with a probability $\beta$ from the edges of the original graph $G = (V, E)$. Therefore, we can assume that each connected component, $CC_i$, on $g_\beta$ is an influence path under the $SIR$ dynamics started from one of the nodes of that connected component, let us say $v$. In other words, $|CC_i| - 1$ shows the number of nodes influenced by $v$. This is also equal to the number of nodes reachable from $v$ on $g_\beta$ as the connected components are mutually disconnected and each node appears in at most one connected component. ∎

**Definition 2 ($G_\beta^v$):** Let's define $G_\beta^v$ as the set of all connected distinct $\beta$-graphs defined on a directed graph $G = (V, E)$, where $\deg_{g_\beta}^{in}(v) = 0$, and $\deg_{g_\beta}^{out}(v) \geq 1$ for all $g_\beta \in G_\beta^v$.

**Lemma 2.** Define $I^{SIR}(v, u, m)$ as the probability of influence of node $v$ on node $u$ under $SIR$ dynamics over all paths of length $m$ in a Directed Acyclic Graph (DAG) $G$. Let us also define $r^{G_\beta^v}(v, u, m)$ as the probability of reachability of $u$ from $v$ with pathes of length $m$ in $G_\beta^v$. Then we have $I^{SIR}(v, u, m) = r^{G_\beta^v}(v, u, m)$.

**Proof.** We employ an inductive approach to prove this lemma. It is clear that the lemma holds for paths of length one, and $I^{SIR}(v, u, 1) = r^{G_\beta^v}(v, u, 1)$ is equal to the probability of existence of a link between $v, u$. Let us assume that the lemma holds for paths of length $m - 1$. To show that it generally holds, we assume that the $m + 1^{\text{th}}$ node ($v_{m+1}$) is the node with the highest rank ($h$) in the topological sorting of $G$. This assumption implies that $v_{m+1}$ does not appear neither in any propagation paths starting from $v_1$ to other nodes, nor in any reachability path starting from $v_1$ and ending in nodes other than $v_{m+1}$. In other words, it means that the reachability and influence probabilities of other nodes is independent from $v_{m+1}$. Given this, the influence probability of node $v_{m+1}$ is the probablity that $v_{m+1}$ becomes influenced by any of its neighbours that have rank $h - 1$ in the topological sort order. This probability is equal to $I^{SIR}(v_1, v_{m+1}, m) = \sum_{u \in \Gamma^{G^T}(v_{m+1})} \beta I^{SIR}(v_1, u, m - 1)$, where $G^T$ shows the transpose of the graph $G$. Similarly, the probability for reachability of $v_{m+1}$ is equal to the probability that an edge exists between $v_{m+1}$ and one of its neighbours at rank $h - 1$ that is already reachable from $v_1$. We can write $r^{G_\beta^v}(v_1, v_{m+1}, m) =$ $\sum_{u \in \Gamma^{G^T}(v_{m+1})} \beta r^{G_\beta^v}(v_1, u, m-1)$. According to the step assumption, we already know that the equality $I^{SIR}(v_1, u, m-1) = r^{G_\beta^v}(v_1, u, m-1)$ holds for any node $u$ on paths of length $m-1$. We see that the equality also holds for $m$, if we substitute $I^{SIR}$ with $r^{G_\beta^v}$ for paths of length $m-1$ in the equation for paths of length $m$, and the proof is complete. ∎

**Theorem 1.** Let us take $I^{SIR}(v, G)$ as the theoretical influentiality of node v in a *directed acyclic graph (DAG) G* over the set of all paths generated by the underlying dynamics, *SIR*. Let's also use $R^{G_\beta^v}(v, G)$ *to denote the theoretical reachability of v in all connected distinct β–graphs, $G_\beta^v$*. Then, we have:

$$I^{SIR}(v, G) = R^{G_\beta^v}(v, G)$$

We prove this theorem by two different methods.

**Proof 1.** In the first method, we define theoretical influentiality of a node as the sum of the influence of that node on each individual node in the original network. Since influence-spread under *SIR* dynamics is a stochastic process, we sum over the expected influence of the spreader node ($v$) on the target node. We define the expected influence to be the probability of influence times the number of nodes becoming influenced. For a target node $u$ (size of the nodes is 1), this value is equal to the probability of influence of $v$ on $u$ over paths of different lengths, or:

$$I^{SIR}(v, u) = \sum_m I^{SIR}(v, u, m) \times 1$$

By summing over all target nodes, we can define the theoretical influence-spread of $v$ as follows:

$$I^{SIR}(v, G) = \sum_u I^{SIR}(v, u) = \sum_u \sum_m I^{SIR}(v, u, m)$$

According to lemma 2, $I^{SIR}(v, u, m) = r^{G_\beta^v}(v, u, m) \forall u, m$, therefore we can rewrite the above equation as:

$$I^{SIR}(v, G) = \sum_u \sum_m r^{G_\beta^v}(v, u, m)$$

According to the definition, $R^{G_\beta^v}(v, G)$ is the theoretical reachability from $v$ that is equal to the aggregated expected reachability for all nodes from $v$ over all random graphs. $r^{G_\beta^v}(v, u, m)$ is defined as the reachability probability of $u$ from $v$ through paths of length $m$ over all random graphs in $G_\beta^v$. To express $R^{G_\beta^v}(v, G)$ using $r^{G_\beta^v}(v, u, m)$, we need to aggregate over all nodes and paths of all lengths. We can write $R^{G_\beta^v}(v, G) = \sum_u \sum_m r^{G_\beta^v}(v, u, m)$, making $R^{G_\beta^v}(v, G)$ and $I^{SIR}(v, G)$ equal and the proof is complete.∎

**Proof 2**. In the second method, we start from $R^{G_\beta^v}(v, G)$. We define the theoretical reachability on $G_\beta^v$ as the expected reachability on all $g_\beta \in G_\beta^v$, we can write:

$$R^{G_\beta^v}(v, G) = \sum_{g_\beta \in G_\beta^v} P(g_\beta) R(v, g_\beta)$$

where $R(v, g_\beta)$ shows the number of nodes reachable from $v$ on $g_\beta$, and $P(g_\beta)$ shows the probability of $g_\beta$. According to Lemma 1, $R(v, g_\beta) = I^{SIR}(v, g_\beta)$, therefore, we can write:

$$R^{G_\beta^v}(v, G) = \sum_{g_\beta \in G_\beta^v} P(g_\beta) I^{SIR}(v, g_\beta)$$

Since the input graph is a DAG, each $g_\beta \in G_\beta^v$ corresponds to an influence path under the SIR dynamics staring from $v$. Therefore, we can substitute $g_\beta \in G_\beta^v$ with a SIR path starting from $v$, denoted by $\pi_v^{SIR}$, and $G_\beta^v$ with the set of all paths under SIR starting from $v$, denoted by $\Pi_v^{SIR}$. We can then rewrite the above equation as follows

$$R^{G_\beta^v}(v, G) = \sum_{\pi \in \Pi_v^{SIR}} P(\pi_v^{SIR}) I^{SIR}(v, \pi_v^{SIR})$$

In the above equation, the right-hand side is actually the theoretical influence-spread of node $v$ in the given graph $G$ under the SIR dynamics and the proof is complete. ∎

Theorem 1 suggests that computing reachability over $\beta$-graphs is an alternative for computing influentiality under *SIR* dynamics. One advantage for computing reachability over computing influence-spread is that it can be done in a non-iterative way, as the edges of the $\beta$-graph are independent and can be generated simultaneously. We use this insight to design a randomized algorithm that estimates influentiality through random generation of $\beta$-graphs.

To capture the aggregated reachability of nodes on a set of $\beta$-graphs ($G_\beta^v$) and accurately approximate the theoretical reachability, inspired by the Reverse-Influence-Sampling (RIS) framework [48], we build a hyper-graph with hyper-edges that represent reachability relation between nodes in the original graph. We then provide a theoretical guarantee that shows a factor of the degree of the nodes in the hyper-graph is a $(1-\epsilon)$ approximation of the theoretical influentiality of the nodes in the original network if enough $\beta$-graphs are sampled. Since *Theorem 1* is stated for a special set of $\beta$-graphs that are connected and distinct ($G_\beta^v$), we make some changes to the straight-forward way of generating $\beta$-graphs to keep everything close to the theory to the best extent possible. In this regard, we define a threshold ($T$) for size of the connected components generated in a $\beta$-graph, and only keep the connected components that have a size greater than or equal to this threshold. This threshold directly depends on the underlying dynamics and its parameters and helps the $\beta$-graph sampling method to generate connected components that represent influence paths under the given dynamics. To illustrate, a random $\beta$-graph might have many connected components of size one or two, while under a certain dynamic with a high influence probability, influence paths of length zero or one are very rare. In such cases, the threshold can help in keeping only the connected components that have an influence path of reasonable length (greater than 1). The threshold ($T$) is a tuneable parameter and is calculated through experiments. Figure 1 shows a visual illustration.

---

**Algorithm 1.** *Randomized Influence Paths Selection* (*RIPS*)
**Input**: A directed graph $G = (V, E)$, $\beta$, a threshold $T$ that depends on $G$ and $\beta$.
**Output**: A ranking of nodes in $G$ based on their influentiality using the aggregated weights of their edges in $G_h$.
1:      For $\theta'$ rounds:
2:          Build a $\beta$-graph $g_\beta$ by selecting $m$ edges each with probability $\beta$ from $E$.
3:          $G_\beta \cup= \{|CC_i| > T : \forall\, CC_i \in g_\beta\}$     # $CC_i$ is a connected component.

4:      $\forall\, CC$ in $G_\beta$:     # Adding one hyper-edge per connected component in $G_\beta$.
5:          $if\ uniform\text{-}weight$:
6:              $W_H(u) = W_H(u) + 1 \quad \forall\, u \in CC$     # Updating nodes degree assuming an unweighted hypergraph.
7:          $else$:
8:              $W_H(u) = W_H(u) + |CC|\beta \deg_G(u) \quad \forall\, u \in CC$     # Updating nodes degree in assuming a weighted hypergraph.

9:      Return $\{(u_1, u_2, \ldots, u_n) : W_H(u_i) \geq W_H(u_j)\ if\ i < j\}$     # Rank nodes based on their degree in the hypergraph.

---

In Algorithm 1, we use a collection of random $\beta$-graphs to approximately replicate $G_\beta^v$ as stated in Definition 2. We use the notation $G_\beta$ for this collection. $G_\beta$ is used in our algorithm to compute reachability from each node. Algorithm 1, or the *Randomized Influence Paths Selection* (*RIPS*), shows the pseudocode of our proposed method.

In the above algorithm, $H = (V_h, E_h)$ is used to denote a hyper-graph where each hyper-edge $e_h \in E$ can link up to $n = |V|$ nodes together, and $G = (V, E)$ is the original network. The hyper-edges in our algorithm are sampled from the original graph. We also propose two versions of *RIPS*, one with uniform weighting, and the other with customized weighting for the hyper edges. In the weighted version, the weights are assigned to differentiate between the nodes based on the number of nodes reachable from them. The weight of a hyper-edge is not equal for all the nodes in that hyper-edge. In our algorithm, the size of a connected component shows the number of nodes reachable from the spreader node in that component. By assigning a weight equal to the size of the connected component, we make sure that a node that can reach more nodes is getting more weight in the hyper-graph compared to a node that only reaches few other nodes in the $\beta$-graph. Moreover, to differentiate between the nodes in the same connected component, we also incorporate the likelihood of each node for being in that connected component. For simplicity, we assume that each node can appear in a given connected component through each of its links. With this assumption, the chance that a node will be in a connected component is $\beta \deg_G u$, or the influence probability times the degree of that node. Therefore, the weight of a hyper-link for the nodes of a connected component is equal to the chance that a node will be in a connected component times the size of the connected component, or $|CC|\beta \deg_G u$, with $|CC|$ being the size of the connected component.

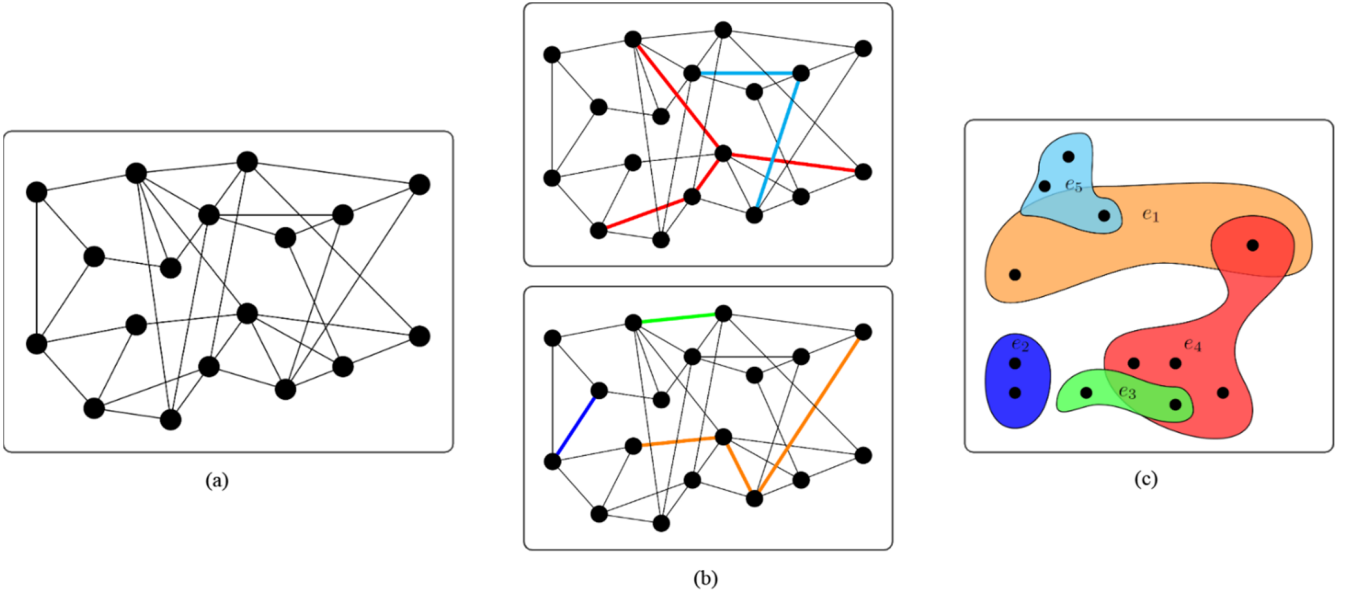

Figure 1. A visual illustration of the RIPS algorithm. Part (a) shows the initial network. Part (b) Shows two different iteration of the randomized sampling of $\beta$-graphs in the RIPS algorithm. Two sampled $\beta$-graph is shown, with 2 and 3 connected components. Part (c) shows the final hyper-graph where a hyper-edge is created for each connected component in the sampled $\beta$-graphs. Finally, the nodes are sorted based on their degree in the hyper-graph.

Inspired by the theoretical analysis conducted by Borgs et al. [50], Nguyen et al. [49], and Tang et al. [50], in the rest of the method section we will theoretically argue how the hyper-graph is useful for computing an approximate solution of the theoretical influentiality of nodes.

**Lemma 4.** The expected influentiality of a spreader node $v$ in a random beta-graph $g_\beta$, can be calculated through the probability that a random node $u$ belongs to the hyper-edge that includes $v$, or
$$E_{g_\beta \sim G}\left(I_{g_\beta}(v)\right) = n \Pr_{g_\beta \sim G}\left(u \in e_H(v)\right)$$

**Proof.** We use $E_{g_\beta \sim G}\left(I_{g_\beta}(v)\right)$ to denote the expected influentiality which is the expected number of nodes that will become influenced if we start the propagation process from node $v$. By dividing this number by $n$ we have the probability of influence of $v$ on a random node $u$ in $G$. According to lemma 2, the probability of influence of $v$ on $u$ is equal to the probability of reachability of $v$ to $u$. Therefore, we can write:

$$\frac{E_{g_\beta \sim G}\left(I_{g_\beta}(v)\right)}{n} = \Pr_{g_\beta \sim G}\left(u \in R(v, g_\beta)\right)$$

All the nodes that are reachable from $v$ on $g_\beta$ constitute a connected component that is also represented by a hyper-edge in *RIPS*. Therefore, we can substitute $R(v, g_\beta)$ with $e_H(v)$ in the above equality. We have:

$$\frac{E_{g_\beta \sim G}\left(I_{g_\beta}(v)\right)}{n} = \Pr_{g_\beta \sim G}\left(u \in e_H(v)\right) \blacksquare$$

**Corollary 1.** $E(I(v)) = n \Pr(u \in E_H(v))$.

**Proof.** This corollary is a generalization to lemma 4. Assuming that the hyper-graph $H$ is built by a sampling a number of $\beta$-graphs, the corollary is yielded by marginalizing over the sampled $\beta$-graphs. Let us denote the set of sampled $\beta$-graphs by $G_\beta$. Lemma 4 holds for every $g_\beta$ in $G_\beta$. If we take a sum over all $g_\beta$ in $G_\beta$ from both sides of the equation, the equality should still hold. According to lemma 4, we have:

$$\frac{E_{g_\beta \sim G}\left(I_{g_\beta}(v)\right)}{n} = \Pr_{g_\beta \sim G}\left(u \in R(v, g_\beta)\right)$$

By summing over $g_\beta$, we can write:

$$\frac{\sum_{g_\beta \in G_\beta} E_{g_\beta \sim G}\left(I_{g_\beta}(v)\right)}{n} = \sum_{g_\beta \in G_\beta} \Pr_{g_\beta \sim G}\left(u \in R(v, g_\beta)\right)$$

$$= \sum_{g_\beta \in G_\beta} \Pr_{g_\beta \sim G}\left(u \in e_H(v)\right)$$

The righthand side of the above equation can also be expressed by $\Pr(u \in E_H(v))$, where $E_H(v)$ refers to the set of all hyperedges in $G_\beta$ that include $v$. By using $E(I(v))$ to represent the sum of the expected influentiality over all $\beta$-graphs, the corollary holds, and the proof is complete.
∎

**Corollary 2.** $E(I(v)) = n\, E(\frac{\deg_H(v)}{m_H})$, with $m_H$ being the number of hyper-edges in the hyper-graph.

**Proof.** This corollary is yielded by expressing the right-hand side of Corollary 1 in terms of the parameters of the hyper-graph $H$. The probability that a random node $u$ belongs to $E_H(v)$, the set of all hyper-edges that include $v$, is equal to the number of hyper-edges that have $v$, or $\deg_H(v)$, over the total number of hyper-edges that exist in the hyper-graph. Since the hyper-graph $H$ is built through random sampling of $\beta$-graphs, we use the expected value of $\frac{\deg_H(v)}{m_H}$ to show the probability of a random node $u$ belonging to $E_H$.
∎

**Theorem 2.** By using $I^{SIR}(v, G)$ to denote the theoretical influentiality of $v$, the following inequality holds with probability $1 - n^{-k}$ for every node $v \in G$ that satisfies $E(I(v)) \leq I^{SIR}(v, G)$ when $\theta \geq \frac{\log 2 + k \log n}{I^{SIR}(v,G)\, \epsilon^2}(8 + 2\epsilon)n$.

$$\left| n \cdot \frac{\deg_H(v)}{m_H} - E(I(v)) \right| \leq \frac{\epsilon}{2} I^{SIR}(v, G)$$

**Proof.** Let's use $\phi$ to denote the probability that node $v$ influences a random node $u$. According to corollary 2 we have:

$$\phi = \frac{E(I(v))}{n} = E(\frac{\deg_H(v)}{m_H})$$

Let's use the Chernoff bounds for the reverse inequality. We have:

$$\Pr\left[\left|n.\frac{\deg_H(v)}{m_H} - E(I(v))\right| \geq \frac{\epsilon}{2}I^{SIR}(v,G)\right] =^{\times \theta}_{/n}$$

$$\Pr\left[\left|\theta.\frac{\deg_H(v)}{m_H} - \theta\phi\right| \geq \frac{\epsilon\theta}{2n}I^{SIR}(v,G)\right] =^{\times\phi}_{/\phi}$$

$$\Pr\left[\left|\theta.\frac{\deg_H(v)}{m_H} - \theta\phi\right| \geq \frac{\epsilon I^{SIR}(v,G)}{2n\phi}\theta\phi\right]$$

As discussed earlier, $\frac{\deg_H(v)}{m_H}$ shows the number of hyper–edges that contain node $v$ over the total number hyper–edges that $v$ could participated in. Therefore, we can say that $\theta.\frac{\deg_H(v)}{m_H}$ is the sum of $\theta$ i.i.d. Bernoulli variables with a mean of $\phi$. According to the assumption we know $\phi \leq E(I(v))/n \leq I^{SIR}(v,G)/n$. We also set $\delta = \frac{\epsilon I^{SIR}(v,G)}{2n\phi}$. According to the Chernoff bounds we have:

$$\Pr\left[\left|\theta.\frac{\deg_H(v)}{m_H} - \phi\theta\right| \geq \delta\theta\phi\right] \leq \exp\left(-\frac{\delta^2}{2+\delta}.\theta\phi\right) \leq^{expand\,\delta}$$

$$\leq \exp\left(-\frac{\epsilon^2 I^{SIR}(v,G)^2}{8n^2\phi^2 + 2\epsilon n\phi I^{SIR}(v,G)}\theta\phi\right) \leq^{elimination}$$

$$\leq \exp\left(-\frac{\epsilon^2 I^{SIR}(v,G)^2}{8n^2\phi + 2n\epsilon I^{SIR}(v,G)}\theta\right) \leq^{n\phi \leq I^{SIR}(v,G)}$$

$$\leq \exp\left(-\frac{\epsilon^2 I^{SIR}(v,G)^2}{8nI^{SIR}(v,G) + 2\epsilon I^{SIR}(v,G)}\theta\right) \leq^{elimination} \exp\left(-\frac{\epsilon^2 I^{SIR}(v,G)}{8n + 2\epsilon n}\theta\right) \leq \frac{1}{n^k}$$

The right–hand side of the last inequality holds for any $\theta \geq \frac{\log 2 + k \log n}{I^{SIR}(v,G)\,\epsilon^2}(8+2\epsilon)n$ and the proof is complete.

∎

**Theorem 3.** Given an input graph $G = (V,E)$, the *RIPS* algorithm approximates the theoretical influentiality of a given node $v$ with a factor of $(1-\frac{\epsilon}{2})$ with a probability of $(1-n^{-k})\left(\frac{\deg_G(v)}{|CC|}\right)^{d_v}$, where $d_v$ is the minimum degree for node $v$ in a random hyper-graph $H$ that satisfy the inequality $n\frac{d_v}{m} > E(I(v))$, $|CC|$ is size of the connected component in the original Graph $G$ that has $v$, and $\theta \geq \frac{\log 2 + k\log n}{I^{SIR}(v,G)\,\epsilon^2}(8+2\epsilon)n$.

**Proof.** Corollary 2 states that $E(I(v)) = n\,E(\frac{\deg_H(v)}{m_H})$. Given a fixed number of $m_H$ (for simplicity shown with $m$), there exists a finite number of hyper-graphs, $\mathcal{H}^m$, with the same number of hyper-edges but different graph topology. By expanding the right-hand of corollary 2, we can write:

$$E\left(\frac{\deg_H(v)}{m_H}\right) = \sum_{H_k \in \mathcal{H}^m} \Pr(H_k)\deg_{H_k}(v)/m$$

We assume that $d_v$ is the minimum degree for node $v$ in a random hyper-graph $H$ that satisfy the inequality $n\frac{d_v}{m} > E(I(v))$. For every degree $\deg_H(v) \geq d_v$, where the inequality $n\frac{\deg_H(v)}{m} < I^{SIR}(v,G)$ does not hold, we can use Theorem 2 in the following way:

$$E(I(v)) \geq n\frac{\deg_H(v)}{m} - \frac{\epsilon}{2}I^{SIR}(v,G)$$

$$\geq I^{SIR}(v,G) - \frac{\epsilon}{2}I^{SIR}(v,G)$$

$$\geq \left(1-\frac{\epsilon}{2}\right)I^{SIR}(v,G)$$

According to Theorem 2, the main inequality holds with probability $1-n^{-k}$, when $\theta \geq \frac{\log 2 + k\log n}{I^{SIR}(v,G)\,\epsilon^2}(8+2\epsilon)n$. The probability for $n\frac{\deg_H(v)}{m} \geq I^{SIR}(v,G)$ is at most $\left(\frac{\deg_G(v)}{|CC|}\right)^{d_v}$, because $v$ requires at least a degree of $d_v$ in the hyper-graph $H$ to satisfy the inequality, which means it should participate in at least $d_v$ connected components of the sampled $\beta$-graphs, where the probability for participating in a sampled connected component is $\left(\frac{\deg_G(v)}{|CC|}\right)$. Therefore, the theorem is held with a probability of at most $(1-n^{-k})\left(\frac{\deg_G(v)}{|CC|}\right)^{d_v}$, and the proof is complete.

■

We use Theorem 2 to show that the *RIPS* algorithm can theoretically obtain an approximation of the theoretical influentiality of nodes with a factor of $(1 - \epsilon)$. However, calculating the exact value for $\theta$ is challenging as it requires the theoretical influence-spread for every node, which is unknown. Instead, we employ an empirical approach to find the right value for $\theta$. Although Theorem 2 is stated for the unweighted version of the *RIPS* algorithm, our empirical results show that the weighted version yields a better performance, therefore we only report the results of the weighted version. We have tested the *RIPS* algorithm on eight undirected graphs that are widely used in the literature and compared the results with the results of 10 existing methods. The results show that our model achieves state-of-art performance on six graphs and a comparable performance on the other two graphs. A detailed discussion of the experimental results is included in the next section.

## 4. Experimental Results

We have selected eight graph datasets widely used in the literature to evaluate the performance of influential node ranking methods. We have included graph datasets varying in size and internal structures to ensure that this selection covers a wide range of graph datasets of different types. Table 1 details the specifications of the datasets used in the experiments. In this table, $\beta_{th}$ is the epidemic threshold reported for each dataset [23], and $\beta$ is the influence probability that we used in our experiments. According to [23], the influence probability $\beta$, should be greater than the epidemic threshold, $\beta_{th}$, to end up having a reasonable size of influenced people. That is why we have set $\beta$ to be slightly larger than $\beta_{th}$ in Table 1.

Table 1. Details of the datasets used for the experimentation. Datasets with different number of nodes and different number of internal connections are selected.

| Dataset | Number of Nodes | Number of Edges | Max Degree | Average Degree | $\beta_{th}$ | $\beta$ |
|---|---|---|---|---|---|---|
| Dolphins | 62 | 159 | 12 | 5.129 | 0.147 | 0.15 |
| Copperfiled | 112 | 425 | 49 | 7.589 | 0.073 | 0.1 |
| Netscience | 379 | 914 | 34 | 4.823 | 0.125 | 0.15 |
| Elegansa | 453 | 4596 | 639 | 20.291 | 0.006 | 0.01 |
| Euroroad | 1,174 | 1,417 | 10 | 2.414 | 0.333 | 0.35 |
| Hamster | 2,426 | 16,631 | 273 | 13.711 | 0.024 | 0.03 |
| PowerGrid | 4,941 | 6,594 | 19 | 2.669 | 0.258 | 0.3 |
| PGP | 10,680 | 24,316 | 205 | 4.554 | 0.053 | 0.1 |

To highlight the advantages and disadvantages of our proposed model, we compare its results with the results of state-of-the-art models reported in the literature. These models include k-shell decomposition centrality (KS)[15], mixed degree decomposition (MMD) [20], semi-local degree and weighted entropy (CDE) [45], mixed core and semi-local degree and weighted entropy (MCDE) [45], extended neighbourhood coreness centrality (CNC+) [23], k-shell iteration factor (Ks-IF) [41], the dynamics-sensitive centrality (DS)[32], diversity-strength ranking (DSR) [46], extended diversity-strength ranking (EDSR) [46], entropy based ranking measure (CRM) [42], and extended cluster coefficient ranking measure (ECRM) [42], which we briefly introduced in section 2. We evaluate the performance of our proposed method using the correlation of the generated ranking with the ground-truth ranking, the uniqueness of ranking, and the computational complexity of the method, each of which are elaborated hereunder.

| Dataset | KS | MMD | CNC+ | KS-IF | MCDE | DS | DSR | EDSR | CRM | ECRM | *RIPS* |
|---|---|---|---|---|---|---|---|---|---|---|---|
| Dolphins | 0.7957 | 0.7812 | 0.8619 | 0.841 | 0.8321 | 0.9085 | 0.8995 | 0.8942 | 0.8584 | 0.9249 | **0.9426** |
| Copperfiled | 0.8691 | 0.8206 | 0.9125 | 0.8725 | 0.8604 | 0.9310 | 0.9299 | 0.9237 | 0.8301 | 0.9273 | **0.9330** |
| Netscience | 0.5797 | 0.5606 | 0.6966 | 0.8364 | 0.6183 | 0.8931 | 0.8243 | 0.8641 | 0.8768 | **0.9006** | 0.8971 |
| Elegansa | 0.7683 | 0.7501 | 0.8109 | 0.6962 | 0.7564 | 0.1330 | 0.8172 | 0.8516 | 0.6595 | 0.7989 | **0.8614** |
| Euroroad | 0.602 | 0.5998 | 0.7188 | 0.8071 | 0.6385 | 0.8473 | 0.7946 | 0.8422 | 0.8646 | 0.8582 | **0.8818** |
| Hamster | 0.7165 | 0.702 | 0.8193 | 0.8787 | 0.725 | 0.8051 | 0.8047 | 0.8175 | 0.8339 | 0.8378 | **0.8574** |
| PowerGrid | 0.5456 | 0.5494 | 0.6564 | 0.7668 | 0.5841 | 0.8270 | 0.7262 | 0.7659 | 0.774 | 0.7899 | **0.8395** |
| PGP | 0.4893 | 0.4776 | 0.6733 | 0.6913 | 0.489 | 0.6901 | 0.7199 | **0.7385** | 0.7158 | 0.7316 | 0.7303 |

Table 2. The Kendal-Tau correlation with the ground-truth ranking for the rankings generated by baseline methods and the proposed method.

### 4.1 Ranking Correlation with Ground-Truth

The main criterion that we use for evaluation is the correlation of the ranking generated by the algorithms ($R$) with the ground-truth ranking obtained by numerical simulations. Since the real influentiality of the nodes is unknown, to obtain an approximation of a node's influentiality, we simulate the underlying dynamics (in this case SIR) starting from the respective node. The number of iterations that we run the SIR simulation varies from dataset to dataset. For each dataset, we set this to be the smallest number that can achieve a unique ranking, whereby nodes with different influential capabilities have different ranks. The approximate influentiality obtained by simulating SIR is used to build the ground-truth ranking list ($GTR$). In this list, the node with the highest influentiality is ranked at the top, and the influentiality of each node is measured by the number of nodes that become influenced if the propagation starts from that given node. We then use Kendall's tau correlation coefficient (KT-correlation) [51] to measure the extent to which the generated ranking ($R$) matches with the ground-truth ranking ($GTR$). KT-correlation takes a value between −1 and +1, whereby a larger value indicates a higher degree of correlation. For two lists of size $n$ we have:

$$\tau(GTR, R) = \frac{1}{n \times (n-1)} \times \sum_{i<j} sign((gtr_i - gtr_j)(r_i - r_j))$$

In the above formula, $sign(x)$ is the sign function that takes 1 if $x > 0$, takes $-1$ is $x < 0$, and takes 0 if $x = 0$.

Table 2 shows the KT-correlation of rankings generated by each of the algorithms, including our proposed model, with the ground-truth ranking. For each dataset, the influence probability is set according to Table 1.

The results confirm that for a $\beta$ close to the epidemic threshold, our proposed model outperforms other algorithms for most of the datasets. For two datasets (Netscience and PGP), the proposed model shows very close results to the top-performer. The major advantage of the proposed model is its dynamics sensitivity that enables *RIPS* to generate a different ranking if the underlying dynamics or its parameters change. This characteristic of *RIPS* contrasts to the static nature of the other methods, as these methods generate the ranking only based on the structural properties of the input network and disregard the underlying dynamics.

Figure 2 shows the KT-correlation between the ranking generated by different models and the ground-truth for different values of $\beta$. We see that the KT-correlation for *RIPS* remains relatively flat when compared with the performance of the other state-of-the-art models. This is particularly evident for DS, EDSR and ECRM which suffer from a significant drop for low values of $\beta$. Similar behaviour for dynamics-independent models have already been reported in the literature [42, 46], but it is surprising that the other dynamics-sensitive approach, DS [32], is not comparable to other dynamics-independent methods such as EDSR and ECRM. For Dolphins and Netscience datasets, MMD and MCDE are among the top performing models for low values of $\beta$. This can be explained by the fact when $\beta$ is low, the length of the influence path is short and the first-order neighbours play an important role in the diffusion process. Therefore, methods that only consider the first-order neighbours, such as MDD and MCDE, achieve a good performance. However, as the $\beta$ increases, we can see that the performance of MDD and MCDE declines significantly. The results also imply that as the size of the network increases, the correlation score of all of the methods drop. This is most likely due to the fact that ranking the nodes of a larger network is more challenging. In the case where there is a high chance of having nodes with similar properties (that should be assigned with the same rank), it may be difficult for the algorithms to identify these similar nodes and give them equal ranks. The best performing baseline models such as DS, DSR, EDSR, CRM and ECRM are showing good performances around the epidemic threshold while their performances drop at both ends of the graph. In regard to the dynamics-independent models, this behaviour suggests that these models are tuned to only work for a certain range of $\beta$. For the dynamics-sensitive method, DS, this behaviour suggests that approximating the non-linear propagation process with linear matrix multiplication in a diffusion dynamic such as SIR works only when the influence probability is close to the epidemic threshold. This highlights one of major shortcomings of the DS [48] method.

**4.2 Uniqueness of Ranking**

Uniqueness of ranking is the second criterion that we use for the evaluation of the models. An ideal ranking should assign unique ranks to each of the nodes in the network, when the influentiality of the nodes are different. Monotonicity relation ($M(R)$) [23] is a metric that is widely used in the literature for measuring the uniqueness of ranking. Monotonicity relation for ranking list ($R$) is defined as follows:

$$M(R) = 1 - \left(\frac{\sum_{r \in R} n_r \times (n_r - 1)}{n \times (n - 1)}\right)^2$$

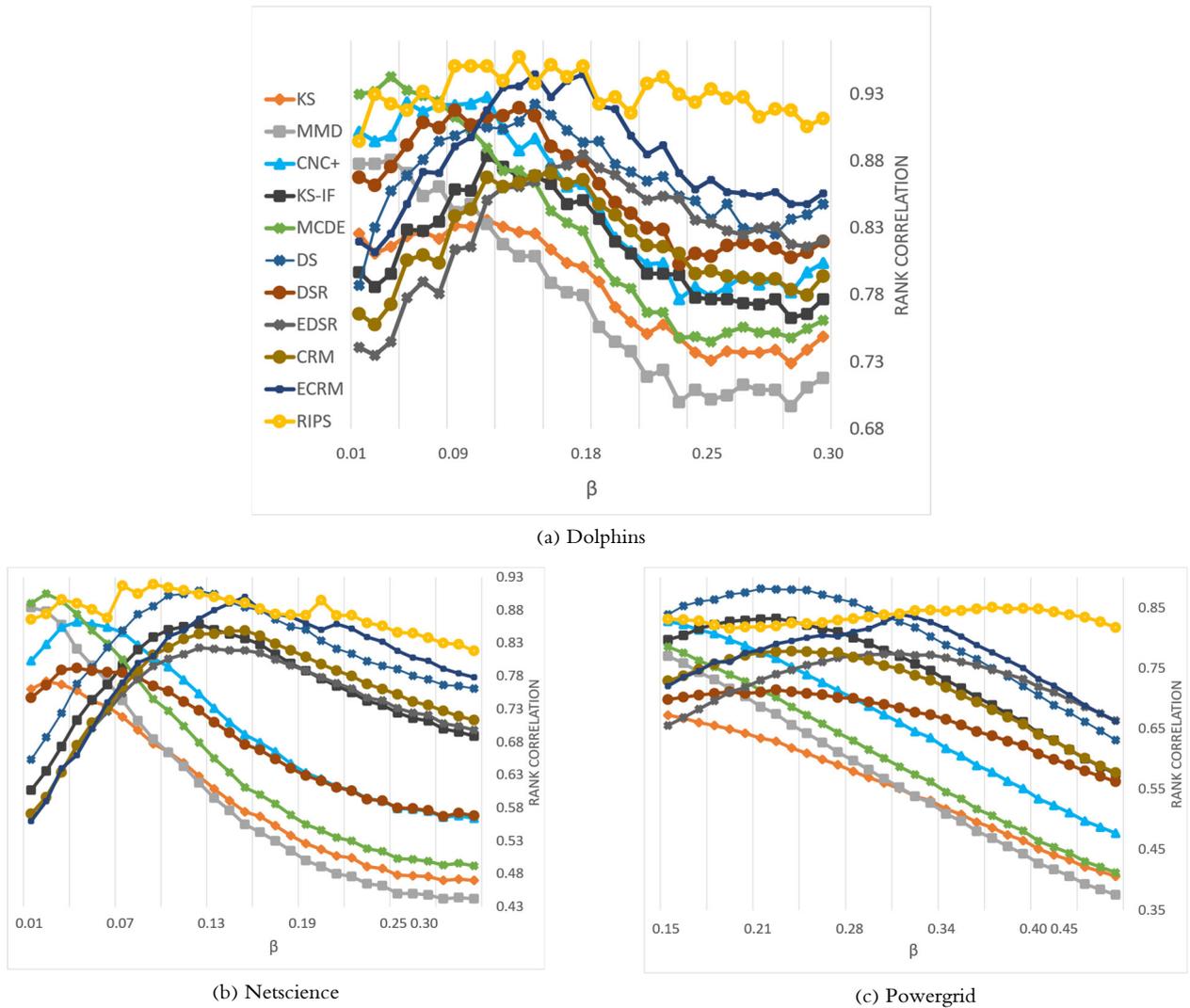

(a) Dolphins

(b) Netscience

(c) Powergrid

Figure 2. The Kendal-tau correlation with the ground-truth ranking for rankings generated by different methods when the influence probability changes.

with $n$ being the number of unique ranks in $R$, and $n_r$ being the number of nodes with the same rank $r$. $M(R)$ gets a value between 0 and 1, whereby a larger value indicates a greater uniqueness and uniformity for the ranking.

Table 3 shows the monotonicity relation values for ranking lists generated by different models. The results show that the *RIPS* algorithm achieves the best performance across all datasets. In fact, the ranking generated by *RIPS* achieves absolute uniqueness for three of the small and medium scale networks, while achieving an almost unique ranking for the remaining datasets. Its lowest performance in terms of ranking uniqueness occurs on the Euroroad dataset which is a not so dense, medium scale network. The reason is that there are many nodes in Euroroad with the same degree and potentially similar structural properties that are hard to distinguish between.

However, we see that the performance of the other dynamics-sensitive method, DS [48], is far from the top performers. Surprisingly, for the Elegansa dataset, the DS method records the lowest score of zero for the uniqueness of ranking, while on the same dataset, the RIPS algorithm scores only a tiny fraction away from absolute uniqueness. The results of table 3 suggest that many nodes are getting the same rank when ranked with a method like DS, while we know that there is another ranking (RIPS ranking) with even higher correlation with the ground truth that assigns almost unique ranks to these nodes. It also highlights yet another drawback of matrix-based methods such as DS. In such methods, the adjacency vector of the nodes plays a key role in approximating the influentiality. However, in a large network, there may be many nodes with similar adjacency vectors (first-order neighbours), that do not have similar second-order and higher-order neighbours. Matrix-based methods such as DS treat the nodes with similar adjacency vectors almost similarly and this creates the uniqueness problem.

To visualise the difference between the rankings generated by a method like RIPS, that generates almost unique ranking, with a method like DS that has a low $MR$ score, we plot the distribution of nodes at different ranks.

Figure 3 shows the percentage of nodes at different ranks starting from rank 1, the most influential node, to rank $N$, where $N$ is the number of nodes in that graph. As the results shows, for our proposed method, RIPS, the distribution of nodes at different ranks is uniform, which means a random node has the same chance of being assigned to a low rank or a high rank, although for the RIPS method unlike other methods, the chance of assigning a node to a high rank is slightly higher. On the other hand, the distribution of nodes at ranks for other methods except DSR is far from being uniform. In fact, these methods are biased toward assigning a very low rank to a random node. Among these methods, the DS method and MMD method have the worst performance with a spiking node-rank distribution around ranks 1 to 100. This is a characteristic expected from methods that mainly rely on first-order neighbours.

Table 3. The $M(R)$ measure for rankings generated by different methods for various datasets. The proposed method outperforms every other model. The ECRM that has the best correlation with ground-truth among the baselines is being outperformed by the DSR and EDSR methods.

| Datasets | KS | MMD | CNC+ | KS–IF | MCDE | DS | DSR | EDSR | CRM | ECRM | **RIPS** |
|---|---|---|---|---|---|---|---|---|---|---|---|
| Dolphins | 0.6538 | 0.8584 | 0.9499 | 0.9842 | 0.9365 | 0.7448 | 0.9979 | 0.9979 | 0.9979 | 0.9969 | **1.0** |
| Copperfiled | 0.7358 | 0.8988 | 0.9792 | 0.9977 | 0.9555 | 0.8279 | 0.9997 | 0.9997 | 0.9997 | 0.9997 | **1.0** |
| Netscience | 0.6643 | 0.7925 | 0.943 | 0.9895 | 0.9034 | 0.7350 | 0.9955 | 0.9951 | 0.9953 | 0.9952 | **1.0** |
| Elegansa | 0.715 | 0.8117 | 0.9765 | 0.9968 | 0.9486 | 0.0000 | 0.9990 | 0.9991 | 0.9989 | 0.9989 | **0.9999** |
| Euroroad | 0.243 | 0.4442 | 0.6954 | 0.9099 | 0.6852 | 0.7355 | 0.9921 | 0.9959 | 0.9978 | 0.9979 | **0.9986** |
| Hamsterster | 0.8779 | 0.8895 | 0.9756 | 0.9851 | 0.9544 | 0.4825 | 0.9544 | 0.9547 | 0.9848 | 0.9858 | **0.9993** |
| PowerGrid | 0.3614 | 0.5927 | 0.7997 | 0.9517 | 0.7566 | 0.7301 | 0.9994 | 0.9999 | 0.9999 | 0.9999 | **0.9999** |
| PGP | 0.5093 | 0.6193 | 0.9032 | 0.981 | 0.6856 | 0.4312 | 0.9994 | 0.9997 | 0.9997 | 0.9997 | **0.9998** |

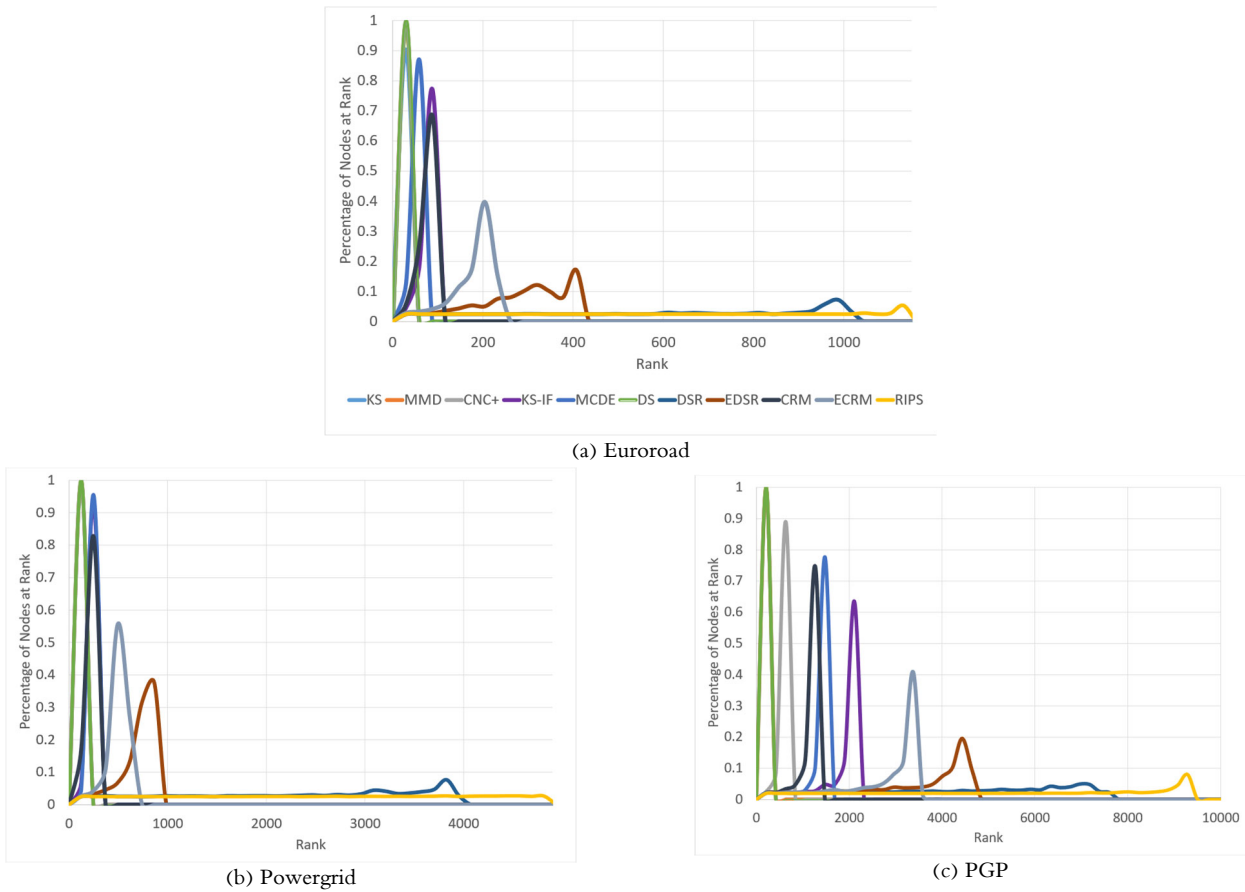

Figure 3. Distribution of nodes ranking generated by different methods for three datasets; Euroroad, Powergrid, and PGP. A uniform distribution implies that there is an equal number of nodes at each rank. RIPS method has the closest distribution to uniform among all methods. Most of the baselines assign a majority of the nodes to the same rank.

Figure 3 shows that methods such as DS and MMD assign no rank greater than 100 to any of the nodes. Moreover, other methods such as KS, KS-IF and CNC+ that only rely on shell-index will also find assigning unique ranks difficult. It is because networks with a lower average degree are divided into less shells when performing a shell-decomposition over the graph. For a graph with around 1100 nodes, the highest rank that these methods assign to the nodes is 80, which means a majority of the nodes share the same rank somewhere between rank $10^{th}$ and $50^{th}$. KS-IF performed much better than KS, MMD and Cnc+ as it uses iterations over the k-shell algorithm. MCDE also performs relatively well, and at points even better than KS-IF; this is mainly because MCDE also takes local information into account. Although ECRM was the top-performer among the baselines in achieving a high correlation with the ground-truth, in terms of uniqueness of ranking, it achieves a similar performance to other baselines where a majority of the nodes are getting the same rank, while there is no node with a rank close to the size of the network. However, *RIPS* remains the top performer in uniqueness of ranking and assigns unique ranks to almost all nodes.

**4.3 Computational Complexity**

Computational complexity accounts for the order of processing time required for running an algorithm and is usually expressed as a function of the input size to the algorithm. In this study, if we assume an input network of $|V|$ nodes and $|E|$ connections, Table 4 shows the computational complexity of each of the baseline methods along with the proposed method.

| Method | Computational Complexity |
|--------|--------------------------|
| KS | $O(|V|^2)$ |
| MMD | $O(|V|^2)$ |
| CNC+ | $O(|V|^2 + |E|)$ |
| KS-IF | $O(|V|^2 + |E|)$ |
| CDE | $O(|V|^2 + |E|)$ |
| MCDE | $O(|V|^2 + |E|)$ |
| DS | $\theta(|V|^2 + |V|)$ |
| DSR | $O(|V|^2 + |E|)$ |
| EDSR | $O(|V|^2 + 2|E|)$ |
| CRM | $O(|V|^2 + |E|)$ |
| ECRM | $O(|V|^2 + 2|E|)$ |
| *RIPS* | $O(|V|\log|V| + 2|E|)$ |

Table 4. Computational complexity of different methods stated in terms of the size, $|V|$, and density, $|E|$, of the input network.

As it can be seen, a majority of the models have a component of $|V|^2$ in their computational complexity. This factor of $|V|^2$ originates from the k-core decomposition algorithm that is used in these methods. For the DS method, this factor is coming from the matrix multiplication within the DS method. The term $|E|$ in the computational complexity of some of the methods denotes that there is a traversal over all the connections. Such methods use the information of all the neighbours of a node when computing the score for that node. For examples, the KS-IF method adds the KS-score of all the neighbours of a node to compute the KS-IF score. A similar idea is employed in other methods with a factor of $|E|$ in their computational complexity.

Our algorithm is made up of three major steps. In the first step which is $\beta$-graph sampling, a Bernoulli trial with success rate of $\beta$ is accomplished for each edge. The computational complexity of this step is $O(|E|)$. Then the connected components in the resulting $\beta$-graph are identified and hyper-edges corresponding to the connected components are added to the hyper-graph. The computational complexity of this step is $O(V + |E|)$. Finally, nodes are sorted based on the degree in the hyper-graph. The computational complexity of the latter step is $O(|V|\log|V|)$, with all the steps adding up to an overall computational complexity of $O(|V|\log|V| + 2|E|)$. Our proposed method achieves a better computational complexity than the baselines and state-of-the-art methods which either use k-core decomposition or matrix multiplication in part of their algorithm. Our algorithm on the other hand uses a randomized sampling approach. Although *RIPS* has a comparable theoretical computational complexity with other methods, its practical run time depends on the choice of $\theta'$, or the number of sampled $\beta$-graphs. A larger number of sampled $\beta$-graphs helps achieve a better approximation of the theoretical influentiality, however, it demands more computation time.

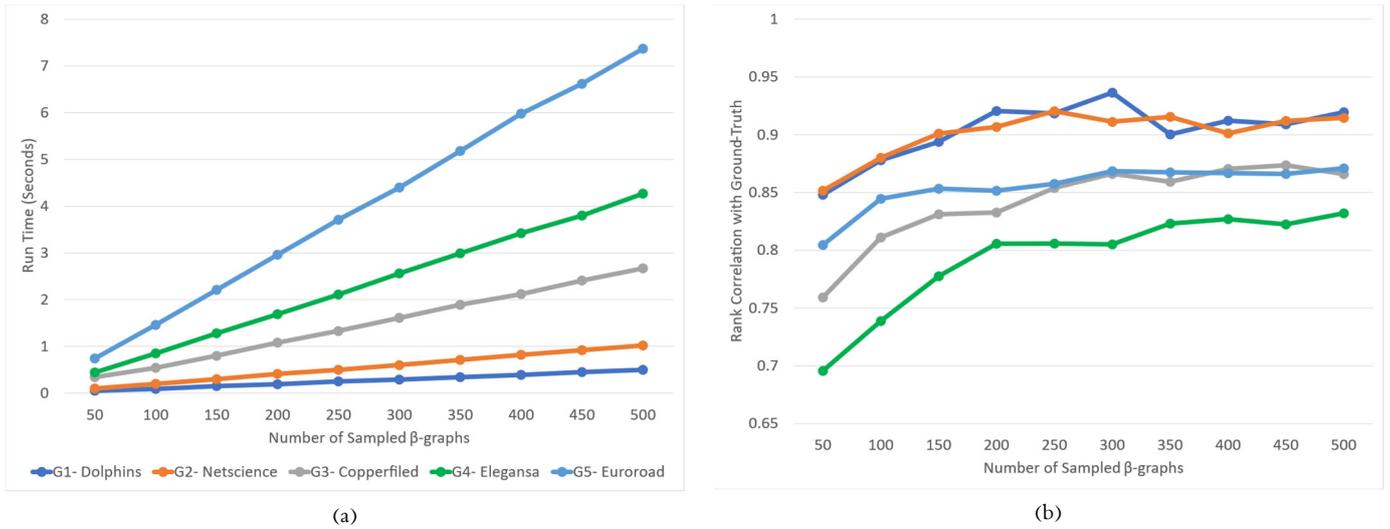

Figure 4. Practical running versus correlation of the generated ranking with the ground-truth ranking for the RIPS method when the number of sampled $\beta$-graphs change.

Figure 4 shows the impact of changing the number of sampled $\beta$-graphs on the running time and correlation with the ground truth for the *RIPS* algorithm. The running time of *RIPS* elevates as the number of nodes in the network, or the number of sampled $\beta$-graphs increase. The change to the correlation with the ground-truth is however slight after the first few iteration, namely 200. These findings suggest that a small number of randomized $\beta$-graphs is sufficient to include every node in the final hyper-graph. By comparing the rank correlations of the *RIPS* as shown in Figure 4-b with the rank correlations of the baselines as of Table 2, we see that a *RIPS* model with 200 sampled $\beta$-graphs outperform the other models or obtains a very close to best performance, while maintaining a very competitive practical running time in comparison to the running time of the baseline methods. With 200 sampled $\beta$-graphs, the *RIPS* algorithm achieves a running time of 1.69 seconds on Eleganza dataset while the *DSR*, *EDSR* and *CRM* methods have a running time of 1.39, 1.30, and 0.58 on this dataset respectively. The results of the practical running time confirm the theoretical analysis of the computational complexity of the models, and also shows that only for a small number of $\beta$-graphs, the *RIPS* algorithm can achieve a state-of-the-art performance in a reasonable time, while the uniqueness of ranking is also maintained.

## 5. Conclusion

With the ever-increasing growth of online social networks, addressing the key questions of the complex networks, such as vital node identification, has attracted much attention. Most of the existing methods for influential node identification and ranking rely only on the structural properties of the network and disregard the underlying diffusion model. One may customize the diffusion models to capture the real-life dynamics in a given complex network, but with the existing structural-based and dynamics-independent vital node identification methods, the impact of changing the diffusion model will not be seen in the generated rankings. In this research, we proposed a randomized dynamics sensitive method, called *RIPS*, for influential node ranking under the *SIR* dynamics. To the best of our knowledge, we are the first to provide a theoretical guarantee for the quality of the solution to the influential node ranking problem. Our results confirm that the ranking of the proposed model changes as the parameters of the underlying dynamics, $\beta$, changes. While the performance of state-of-the-art models drop for low values of $\beta$, the performance of our proposed algorithm stays within a small neighbourhood of its performance on low and high values of $\beta$. Moreover, the proposed model also achieves the best performance in terms of uniqueness of ranking outperforming the state-of-the-art model, *ECRM*, while maintaining a competitive running time. Compared to other existing dynamics-sensitive approach, such as DS, RIPS shows a positive sensitivity to the change of dynamics parameter and its performance stays at the top, while the DS method fails to show a steady performance when the dynamics parameters change. Our proposed algorithm also decisively outperforms DS in terms of the uniqueness of ranking. RIPS uniformly distributes ranks to the nodes, while DS tends to assign only low ranks to the nodes. Future works can investigate the applicability of a dynamics sensitive model on other dynamics such as linear threshold, independent cascade or *SIS*, on different types of networks such as temporal networks.